            \def\fnote#1{\footnote}

\def\be{\begin{equation}}           \def\ee{\end{equation}}
\def\bear*{\begin{eqnarray*}}       \def\eear*{\end{eqnarray*}}
\def\dg{\dagger}                    \def\pr{\prime}
\def\nn{\nonumber}
\def\la{\langle}                    \def\ra{\rangle}
\def\l{\left}                       \def\r{\right}
\def\bc{\begin{center}}             \def\ec{\end{center}}
\def\ni{\noindent}
\documentstyle[12pt]{article}
\begin{document}
\begin{flushright} Preprint INRNE--TH--\,96/\,6 (April 1996)\\
quant-ph/9609017
\end{flushright}
\medskip
\bc
		{\bf SCHR\"{O}DINGER INTELLIGENT STATES AND LINEAR AND
				QUADRATIC AMPLITUDE SQUEEZING}
\ec
\medskip
\centerline{\bf D.~A. Trifonov}
\centerline{Institute of Nuclear Research,}
\centerline{72 Tzarigradsko Chaussee,}
\centerline{1784 Sofia, Bulgaria}
\medskip \medskip

\bc
\begin{minipage}{13cm}
\centerline{\small {\bf Abstract}}
\medskip
{\small A complete set of solutions $|z,u,v\ra_{sa}$ of the eigenvalue
equation $(ua^2+va^{\dg 2})|z,u,v\ra = z|z,u,v\ra$ ($[a,a^\dagger]=1$) are
constructed and discussed.  These and only these states minimize the
Schr\"{o}dinger uncertainty inequality for the squared amplitude (s.a.)
quadratures.  Some general properties of Schr\"{o}dinger intelligent
states (SIS) $|z,u,v\ra$ for any two observables $X, Y$  are discussed,
the sets of even and odd s.a.  SIS $|z,u,v;\pm\ra$ being studded in
greater detail.  The set of s.a.  SIS contain all even and odd coherent
states (CS) of Dodonov, Malkin and Man'ko, the Perelomov $SU(1,1)$ CS and
the squeezed Hermite polynomial states of Bergou, Hillery and Yu.  The
even and odd SIS can exhibit very strong both linear and quadratic
squeezing (even simultaneously) and super- and subpoissonian statistics as
well.  A simple sufficient condition for superpoissonian statistics is
obtained and the diagonalization of the amplitude and s.\,a.  uncertainty
matrices in any pure or mixed state by linear canonical transformations is
proven.
\vspace{0.2cm}

PACS number(s): 03.65.Fd, 42.50.Dv }
\end{minipage}
\ec
\vspace{1cm}
\baselineskip=18pt
\bc
{\large {\bf I. Introduction}}
\ec
\medskip
In the last decade or so there has been a great interest
in squeezed states (SS)\cite{Caves} as nonclassical states with promising
applications (see the review papers\cite{Walls} and
references therein).  Originally SS of electromagnetic field are defined as
states in which the variance of one of the two quadratures $q$
and $p$ of the photon annihilation operator $a$, $a=(q+ip)/\sqrt 2$, are
less than the variance $\Delta_0 = 1/\sqrt{2}$ of $q$ and $p$ in the
Glauber coherent states (CS) $|\alpha\ra$\cite{Glauber}. Such SS are the
Stoler states $|\zeta,\alpha\ra$\cite{Stoler}, which are the
same\cite{D93} as the Yuen two photon CS\cite{Stoler}, the Dodonov et al.
correlated states\cite{DKM} and the CS constructed earlier in
refs.\cite{{MMT},{Holz}}. In analogy to the term canonical
CS\cite{Klauder} for the CS $|\alpha\ra$ one can call the Stoler states
and their equivalents {\it canonical} SS in order to distinguish them from
other types of $q$-$p$ SS\cite{{Kral},{BuzekJeh},{Hach}} and from
SS for other
observables\cite{{Buzek},{WodE},{Hillery},{Peg},{Ueda}}.
The $q$-$p$ SS are also called amplitude SS or ordinary SS.  The reduction
of the variance of an observable with
continuous spectrum (continuous observable)
is of practical interest since there are
no (normalizable) states with vanishing variance of such observables. In
quantum optics the reduction of fluctuations (squeezing) in $q$ or $p$
entails squeezing of magnetic ($\vec{{\cal H}}$) or
electric field ($\vec{{\cal E}}$).  The two quadratures $X_{sa},\,Y_{sa}$
of $a^2$ (the s.\,a. quadratures),
\be      %1
a^2 = \frac{1}{\sqrt 2}(X_{sa} + iY_{sa})
\ee
are continuous. They are related to the two invariant characteristics
$\vec{\cal E}\vec{\cal H}$ and ${\cal E}^2 - {\cal H}^2$ of the field.
In case of mass particle they describe the energy of the inverted
oscillator and the friction respectively.  Therefor it is of interest to
look for the states in which these quantities are subfluctuant, i.e. to
look for s.\,a. SS. Another motivation is the result of paper
\cite{BHYu}: s.\,a. squeezing (called also quadratic squeezing) in a
given field mode can be transferred to another mode as ordinary one.

With the aim to look for squared amplitude SS and for new nonclassical
states we construct in this paper the family of all states which
satisfy the equality in the Schr\"{o}dinger uncertainty relation (u.\,r.)
(called also Robertson-Schr\"{o}dinger
u.\,r.)\cite{{Schroedinger},{Robertson}} for the quadrature components
$X_{sa},\,Y_{sa}$ of the squared boson operator $a^2$. This is achieved by
solving the eigenvalue problem for the complex linear combination $ua^2 +
va^{\dg 2}$ which in this case is the necessary and sufficient
condition for the minimization of Schr\"{o}dinger inequality\cite{D94}.
States which provide the equality in Schr\"{o}dinger relation should be
referred here as Schr\"{o}dinger {\it intelligent states} ( SIS).
Similarly, the states which minimize the Heisenberg u.\,r. will be called
Heisenberg intelligent states (HIS). HIS are a subset of SIS. The
constructed s.\,a. SIS $|z,u,v\ra_{sa}$ turned out to exhibit very strong
linear and quadratic squeezing (even simultaneously), sub- and
superpoissonian photon statistics and to contain in a natural way several
known sets of states (Dodonov, Malkin and Man'ko even and odd CS (e.\,o.
CS)\cite{DMM}, the Perelomov SU(1,1) CS\cite{Perelomov}, the Bergou,
Hillery and Yu squeezed Hermite polynomial states\cite{BHYu}, the
Spiridonov parity CS\cite{Spiridon}). The principle possibility of joint
linear and quadratic amplitude squeezing stems from the fact that
commutators of $q,\,p$ and $X_{sa},\,Y_{sa}$, are not positive (nor
negative) definite (see section IV).

A definition of SS for any pair of (dimensionless) observables $X,\,Y$ has
been given in ref. \cite{WodE} on the base of the Heisenberg  uncertainty
relation (established for arbitrary $X,\,Y$ in fact by
Robertson\cite{Robertson}): a state $|\psi\ra$ is $X$-$Y$ SS if one of the
two squared variances $\Delta^2_X(\psi)$ or $\Delta^2_Y(\psi)$ is less
than one half of the modulus of the mean of the commutator $[X,Y]$,
\be      %2
\Delta^2_i(\psi) <  \frac{1}{2}\,|\la\psi|[X,Y]|\psi\ra|, \qquad i=X \quad
{\rm or}\quad Y.
\ee
This definition has been used in refs.\cite{{Hillery},{Buzek}} to examine
the quadratic squeezing in e.\,o. CS\cite{DMM} and the squeezing of
generators $K_1,\,K_2$ of $SU(1,1)$ in Perelomov $SU(1,1)$ CS
\cite{Perelomov} and in Barut-Girardello CS\cite{BG}.  However in many
cases the inequality in eq.(1) holds when both $\Delta^2_i$ and
$|\la[X,Y]\ra|$ are very large and even tend to infinity and then one
hardly can call the corresponding states squeezed.  Such are the cases of
the $SU(1,1)$ CS with $X,\,-Y$ being the generators of $SU(1,1)$
\cite{Buzek} and the cases of quadratic squeezing, considered
in ref.\cite{Hillery}.

Here we use the more precise definition: a state is $X$-$Y$ SS if
\be      %3
 \Delta_i(\psi)\, < \, \Delta_0,\qquad  i=X \quad{\rm or}\quad Y,
\ee
where $\Delta^2_0$ is the joint minimal value of the two squared variances
$\Delta^2_X(\psi),\, \Delta^2_Y(\psi)$ and half of the mean of the
commutator $|\la\psi|[X,Y]|\psi\ra|/2$: $\Delta^2_0$ is minimal level at
which the equality of the above three quantities,
\be      %4
\Delta^2_X(\psi) = \Delta^2_Y(\psi) = \frac{1}{2}|\la\psi|[X,Y]|\psi\ra|
\ee
 can be
maintained.  This is very natural definition since (as one can easily
verify) the equalities (4) yield the equality in the Heisenberg u.\,r. and
eqs.(4) hold if and only if $\psi$ is an eigenstate of one of the two
nonhermitean operators $X\pm iY$ (see Proposition 1).  Such eigenstates
are denoted here as $|z\ra$, $z$ being the (complex) eigenvalue.  Note
that if a state is SS according to the definition (3) then the inequality
(2) follows, while the inverse is not true. The explicitly considered
s.\,a. and $SU(1,1)$ squeezing in refs.\cite{{Hillery},{BHYu},{Buzek}} (in
canonically squeezed Fock states, in Perelomov CS and in Barut-Girardello
CS) obeys the relative definition (2) only, not (3). We are looking here
for s.\,a. SS which can exhibit strong squeezing according to the
definition (3).

A natural term for states $|z\ra$ which obey eq.(4) is $X$-$Y$ {\it equal
uncertainty} HIS. In case of the canonical variables $q$ and $p$ the equal
uncertainty HIS $|z\ra$ coincide with the canonical CS (the Glauber CS)
$|\alpha\ra$\cite{Glauber} and they are minimum uncertainty states. In
general case however the equal uncertainties depend on some state
parameters, so that the equal uncertainty HIS are not minimum uncertainty
states. The latter states are among the HIS $|z\ra$ and are denoted here
as $|z_0\ra$.  The corresponding variance $\Delta(z_0)$ is denoted in the
definition (3) as $\Delta_0 $.

There has been a resurgence of interest in the last few years in the even
and odd CS (e.\,o. CS)$|\alpha_{\pm}\ra$\cite{{DMM},{Perina}}  as one of
the promising examples of superposition of macroscopically distinguishable
states (Schr\"{o}dinger cat states)(see e.g.
refs.\cite{{Hach},{Spiridon},{GGKnight},{Ansari},{Kuang}}).  These states
can be experimentally realized in several ways
\cite{{Hach},{GGKnight}} and could be used for example in interferometric
gravitational wave detector to increase its sensitivity \cite{Ansari}.  The
nonclassical properties of e.\,o.  CS have been considered in e.g.
ref.\cite{{BuzekJeh},{Vinogradov},{Barranco}}. They can exhibit ordinary
amplitude squeezing (i.e.  $q,\,p$ squeezing) of about 55 \%
\cite{BuzekJeh}, but, like the Perelomov CS (which also are even/odd
states),  do not exhibit quadratic squeezing according to definition (3).
Along these lines our aim is to construct generalized even and odd CS
(denoted here as $|z,u,v;\pm\ra$) which do exhibit strong quadratic
squeezing and contains the latter states as natural subsets.  The set of
these generalized e.\,o. CS are constructed as SIS  and contain also the
"Hermite polynomial amplitude-squared SS" of Bergou, Hillery and
Yu\cite{BHYu} (the latter's being a subset of the HIS).

As we have noted above the method we use is based on the minimization of
the Schr\"odinger u.\,r. and on the result of paper \cite{D94} that a
sufficient condition for a state $|\psi\ra$ to minimize this relation for
quadrature components $X$ and $Y$ of any operator $A$, is
$|\psi\ra$ to be eigenstate $|z,u,v\ra$ of the linear combination
$uA+vA^\dg$ ($u,v$ - complex parameters) of $A$ and $A^\dg$ (see
eq.(6)).  This is a natural extension to any nonhermitean operator $A$ of
the known property of the canonical SS to be eigenstates of the linear
combination $ua+va^\dg$ (Bogolubov transform of $a$ and
$a^\dg$,\,\, $|u|^2-|v|^2=1$) and to minimize the Schr\"odinger u.\,r..
The property of canonical SS to obey the equality in Schr\"odinger relation
for $q$ and $p$ was established in fact in paper \cite{DKM}, where such
states have been called {\it correlated}.  Eigenstates of linear
combinations $ua+va^\dg$ have been constructed and discussed earlier in
refs. \cite{MMT} as time evolution of initial Glauber CS of quadratic
Hamiltonian systems. Here the generalized e.\,o.  CS $|z,u,v;\pm\ra$ are
constructed as two independent solutions of the eigenvalue eq.(37), i.e. they
are e.\,o. square amplitude SIS.  Any s.\,a.  SIS $|z,u,v\ra_{sa}$ is a
linear combination of these e.\,o. SIS. In particular any equal uncertainty
s.\,a. HIS $|z\ra_{sa}$ is a linear combination of the ordinary e.\,o. CS.
The Yurke-Stoler states (and their one angle parameter
generalization\cite{YS}) and the Spiridonov\cite{Spiridon} parity states
are also such square amplitude HIS $|z\ra$. We note that HIS for
the pair $K_1,\,K_3$, i.e. $K_1$-$K_3$ equal uncertainty HIS, ($K_i$
being the generators of $SU(1,1)$) are constructed in the very recent
paper\cite{Puri}.

The paper is organized as follows. In section II we consider a sufficient
condition (eq.(6)) for a state $|\psi\ra$ to minimize the Schr\"odinger
u.\,r. for the quadratures $X,\,Y$ of any operator $A$.  We also prove that
a necessary and sufficient condition for a state $|\psi\ra$ to be $X$-$Y$
equal uncertainty HIS is to be an eigenstate of $A$ or $A^\dg$. In
analogy to the canonical case the general $X$-$Y$ squeeze operator
$S(u,v)$ is defined as a map from equal uncertainty HIS $|z\ra$ to the SIS
$|z,u,v\ra$. The $2\times2$ $X$-$Y$ uncertainty matrix in any state can be
diagonalized by $SU(1,1)$ linear transformations of $X$ and $Y$.\\ In
section III first we review some properties of the ordinary e.\,o.  states
and some other Schr\"odinger cat states in the light of the Heisenberg
u.\,r.  The diagonalization of $3\times3$ and $2\times2$ s.\,a.
uncertainty matrices by linear commutator preserving transformations (in
particular by canonical ones) is considered.  In subsection III.B we
construct the corresponding generalizations as e.\,o.  SIS solving the
eigenvalue equation (37) in the Glauber CS representation and reveal some
important particular cases of these e.\,o. SIS.  Completeness properties of
the e.\,o. SIS and the half unitarity (isometricity) of s.\,a. squeeze
operator are commented.\\ In section IV the nonclassical properties of the
e.\,o.  SIS are considered: some representative results are shown on Figs.
1-3. A simple sufficient condition for superpoissonian statistics is
obtained in terms of the "length" of the quasispin mean vector.  Finally in
Section V we discuss the $sl(2,C)$ algebraic properties of SIS and problems
of stable evolution of and generation of s.\,a. SIS and SS from
known states, in particular by canonical squeezing of finite superpositions
of Fock states and of e.\,o. CS.
% In this paper we use the following abbreviations: CS = coherent states,
% SS = squeezed states, SIS = Schr\"odinger intelligent states, HIS =
% Heisenberg intelligent states, s.\,a.  = squared amplitude, u.\,r. =
% uncertainty relation, e.\,o. = even and odd.
\bigskip
\bc
{\large {\bf II. The Heisenberg and Schr\"{o}dinger intelligent states }}
\ec
\medskip
In this section we consider the minimization conditions of the
Schr\"{o}dinger uncertainty relation (u.\,r.)  for any two observables
$X,\,Y$ and discuss some general properties of the minimizing states
(called here Schr\"{o}dinger intelligent states (SIS)), squeeze operators
and $X$-$Y$ uncertainty matrix.

The  Schr\"odinger u.\,r. \cite{Schroedinger} for any two observables
$X,\,Y$ and any (pure or mixed) state $\rho$ reads
\be      %5
\Delta^2_X(\rho)\Delta^2_Y(\rho) - \Delta^2_{XY}(\rho) \ge
\frac{1}{4}|{\rm Tr}(\rho\,[X,Y])|,
\ee
where $\Delta^2_X(\rho),\,\, \Delta^2_Y(\rho)$ are the squared variances of
$X$ and $Y$ in the state $\rho$ and $\Delta_{XY}(\rho)$ is their
covariance. It recover the Heisenberg (in fact
Robertson--Heisenberg\cite{Robertson}) relation which is obtained with
$\Delta_{XY} = 0$ in eq.(5). The equality in the above relation (5) holds
in pure states $|z,u,v\ra$ {\it only}, which obey the eigenvalue eq.(6),
\be      %6
(uA+vA^\dg)|z,u,v\ra = z|z,u,v\ra,
\ee
where $z$ is the (complex) eigenvalue, $u,\,v$ are arbitrary complex
parameters and
\be      %7
 A=\frac{1}{\sqrt{2}}(X+iY).
\ee
Eq.(6) is a sufficient condition for the equality in eq.(5)\cite{D94}. In
most cases it is also necessary. They are only some of the eigenstates of
$X$ and $Y$, if exist, which make exceptions.  When the two observables
$X$ and $Y$ are continuous (i.e.  with no discrete spectrum of their
eigenvalues) then the commutator $[A,A^\dg] = -i[X,Y] $ is positive or
negative definite and in such cases the eigenvalue eq.(6) is necessary and
sufficient condition for the equality in the Schr\"odinger relation (5)
\cite{D94}.  We call such minimizing states $|z,u,v\ra$ Schr\"{o}dinger
intelligent states (SIS), or more precisely $X$-$Y$ SIS.  The term
intelligent state (IS) is introduced in ref. \cite{Aragone} on the example of
spin states which provide the equality in the Heisenberg u.\,r. (i.e.
$\Delta_{XY}(\psi)=0$ in eq.(5)).  The eigenvalue eq.(6) for the
canonical observables $q,\,p$ has been taken in ref. \cite{DKM} as a
definition of the {\it correlated states}.  Correlated states
coincide\cite{D93} with the canonical SS
$|\zeta,\alpha\ra$\cite{Caves},  (or Stoler states\cite{Stoler}),
\be      %8
|\zeta,\alpha\ra =
\exp\l[\frac{1}{2}(\zeta^*a^2-\zeta\,a^{\dg 2})\r]\,|\alpha\ra
\equiv S(\zeta)|\alpha\ra = S(\zeta)\,D(\alpha)|0\ra,
\ee
which in the Yuen notations\cite{Stoler} are $|\alpha;\mu,\nu\ra$,
$|\mu|^2-|\nu|^2 = 1$.  So the
canonical SS should be referred also as canonical SIS or $q$-$p$ SIS.  The
squeezed vacuum states $|\zeta,0\ra$ coincide with the Perelomov $SU(1,1)$
CS $|\xi;0\ra$ \cite{Perelomov}, $\xi = \tanh|\zeta|\,{\rm
exp}(-i\theta)$, $\theta = {\rm arg}\zeta$ for the nonsquare integrable
representation with Bargman index $k=1/4$.

In any $X$-$Y$ SIS $|z,u,v\ra$ the three second moments of $X$ and $Y$ are
proportional to the mean commutator $[A,A^\dg] = -i[X,Y]$
and read
\be      %9
\Delta^{2}_X(z,u,v)=\frac{1}{2}\frac{|u-v|^2}{|u|^2-|v|^2}\la
[A,A^\dg]\ra,
\ee    \be      %10
\Delta^{2}_Y(z,u,v)=\frac{1}{2}\frac{|u+v|^2}{|u|^2-|v|^2} \la
[A,A^\dg] \ra,
\ee    \be      %11
\Delta_{XY}(z,u,v)=\frac{ 2\,{\rm Im}\,(u^*v)}
{|u|^2-|v|^2}\la[A,A^\dg]\ra.
\ee
where $\la[A,A^\dg]\ra \equiv \la v,u,z|[A,A^\dg]|z,u,v\ra$. One
can check that the above second moments obey the equality in the
Schr\"odinger relation (5) identically.

Formulas (9)-(11) are valid for any $u,\,v$ but the relation $u\,=\,v$ is
admitted only if the commutator $[A,A^\dg]$ is not positive (negative)
definite. In the latter case at $|u|^2-|v|^2=0$ the mean commutator also
vanishes. If the commutator $[A,A^\dg]$ is positive (negative)
definite then normalized SIS $|z,u,v\ra$ exist for $|u|^2-|v|^2 > 0$
($|u|^2-|v|^2 < 0$) only \cite{D94}. In such cases we easily derive from
the eigenvalue equation that the number of significant real parameters of
SIS is four (or two complex). Such are for example $z/u$ and $v/u$. In
case of positive commutator $[A,A^\dg]$ it is also convenient to fix
\be      %12
|u|^2-|v|^2=1
\ee
(or $|u|^2-|v|^2=-1$ in case of negative commutator) as one usually does in
the canonical case to ensure the boson commutation relations for the
transformed operators $a^\pr = ua+ va^\dg$ and
${a^\pr}^\dg = u^{*}a^\dg+ v^{*}a$.  Here we note that for any
operators $A,\,A^\dg$ the relation (12) provides invariance of the
commutator $[A,A^\dg] = -i[X,Y]$ under linear transformation,
\be      %13
A \longrightarrow A^\pr = uA+vA^{\dg},
\ee
where $X^\pr,\,Y^\pr$ are quadratures of $A^\pr$ ($(12) \rightarrow
[A^\pr, A^{\pr \dg}] = [A, A^{\dg}]$),
\be      %14
A^\pr = \frac{1}{\sqrt 2}(X^\pr + iY^\pr).
\ee
We have to warn that in general $A$ and $A^\pr$ are not unitary
equivalent (they are equivalent in the case of $A=a$ due to the Stone - von
Neuman theorem).  The commutator $[A,A^\dg]$ is positive definite for
$A$ being e.g. any positive integer power of the boson destruction operator
$a^k$ or lowering Weyl generator of any $SU(1,1)$ discrete series
representation ${\cal D}^{-}(k)$.
The commutator is not positive nor negative definite
for example in the case of $A = J_{\pm}$, $J_\pm$ being spin operators
($SU(2)$ generators) and of $A = \tilde{q} + i\tilde{ X_{sa}}$, where
$\tilde{q}$ is one of the quadratures of $a$, and $\tilde{ X_{sa}}$ is one
of the quadratures of $a^2$. In the latter cases the SIS $|z,u,v\ra$ exist
for any value of $u$ and $v$, including $u=\pm v$.

If SIS $|z,u,v\ra$ exist for $[A,A^{\dg}]$ positive (negative) definite
then we see from formulas (9)-(11), that $u = 0$ ($v=0$) is not admitted,
that is the eigenstates of $A^{\dg}$ ($A$) do not exist. For positive
(negative) definite commutator the limit $v=0$ ($u=0$) is admitted: then
the covariance of $X$ and $Y$ is vanishing and the two variances
$\Delta^2_X$, $\Delta^2_X$  are equal to each other and to the one half of
the mean commutator $\la[A,A^\dg]\ra$, obeying the equality
in the Heisenberg relation. That is the states $|z,u,v=0\ra$
($|z,u=0,v\ra$) are nonsqueezed $X$-$Y$ HIS, i.e.  $X$-$Y$ equal
uncertainty HIS.  For equal uncertainty HIS one can prove the following

{\it Proposition 1}. A state $|\psi\ra$ is equal uncertainty HIS for given
two physical observables (hermitian operators) $X,\,Y$ {\it if and only
if} it is an eigenstate $|z\ra$ of $A \equiv (1/\sqrt{2})(X+iY)$ or
$A^\dg$,
\be      %15
A|z\ra = z|z\ra, \quad {\rm or} \quad A^\dg|z\ra = z|z\ra.
\ee
The proof can be carried out considering the means of the operators
$F^\dg F$ and $FF^\dg$, where $F=X +iY-\la X +iY\ra$, and using
positivity of the norm of Hilbert space vectors. \\
Recall that the eigenstates $|z\ra$ are known in the following cases of
$A$: $A=a$ (Glauber CS $|\alpha\ra$); $A=a^2$ (e.\,o. CS
$|\alpha_{\pm}\ra$\cite{DMM}); $A=a^k,\, k \ge 2$ ("$k$-photon
CS"\cite{{BuzekJeh},{Kral}}); $A=K_{\mp}$ (Barut-Girardello CS\cite{BG}),
$K_{\mp}$ being the lowering and raising generators of square integrable
(discrete series) representations ${\cal D}^{\pm}(k)$ of $SU(1,1)$ ;
$A=J_\pm$, $J_\pm$ being the generators of $SU(2)$ (spin operators). In all
cases but $A=a^k$, $k\ge2$, the generalizations of the equal uncertainty
HIS $|z\ra$ to the SIS $|z,u,v\ra$ are established \cite{{DKM},{D94}}.
In the next section we shall construct the SIS $|z,u,v\ra$  for the case
$A=a^2$, generalizing in this way the e.\,o.  CS $|\alpha_{\pm}\ra$.  But
before this let us consider the $X$-$Y$ uncertainty matrix and the notion
of squeeze operator in the light of Heisenberg and Schr\"odinger u.\,r..

The linearity of the transformation $A \rightarrow A^{\pr}$ entails
linear relations between second moments of the new quadratures
$X^\pr,\,Y^\pr$ and second moments of the old ones $(X, Y)$ in any
pure or mixed state $\rho$.  It turned out that the constrain (12), which
makes the commutator preserving transformation (13) belonging to $SU(1,1)$,
ensure the invariance of the determinant of the uncertainty matrix
$\sigma(\rho)$. Moreover, the linear $SU(1,1)$ transformation can be used
to diagonalize this uncertainty matrix for any pure or mixed state $\rho$.
We formulate this statement as a second proposition,

{\it Proposition 2}. The $SU(1,1)$ linear transformation of any pair of
operators $A=(X+iY)/\sqrt{2}$ and
$A^\dg$ preserves the determinant of the $X$-$Y$ uncertainty matrix
$\sigma(\rho)$ invariant and can diagonalize $\sigma(\rho)$ for any pure or
mixed state $\rho$.

Proof. The proof uses significantly the transformation property of the
uncertainty matrix $\sigma(\rho;X,Y)$,
\be        %16
\sigma_{ij}(\rho) = \Delta_{ij}(\rho),\quad i=X, Y,\,\, j=X, Y.
\ee
under linear transformations (13). The new quadratures
$X^\pr,\,Y^\pr$ are related to the old one by means of the
symplectic matrix $\Lambda \in Sp(2;R)$
\be  %17
\left(\begin{array}{c} X^{\pr} \\ Y^{\pr} \end{array}\right) =
\Lambda(u,v)\left(\begin{array}{c} X \\ Y\end{array}\right),
\quad \Lambda =
\left(\begin{array}{lr}{\rm Re}(u+v) & -{\rm Im}(u-v)\\
                       {\rm Im}(u+v) & {\rm Re}(u-v)
\end{array}\right).
\ee
Then the uncertainty matrices for the new and old quadratures
are $\Lambda$ congruent
\be      %18
\sigma^{\pr} \equiv \sigma(\rho;X^\pr,Y^\pr) =
\Lambda(u,v)\sigma(\rho;X,Y)\Lambda^{T}(u,v),
\ee
wherefrom it follows that ${\rm det}\sigma^{\pr} = {\rm det}\sigma$.

The possibility to diagonalize the uncertainty matrix by means of
symplectic matrix $\Lambda$ (corresponding to linear transformation (13)) is
based on its positive definiteness and on its positive
determinant\cite{Gantmaher}. The positivity of ${\rm det}\sigma$ follows from
the Schr\"odinger relation
(5). So we have to prove the positive definiteness of $\sigma$ only. For
this purpose we have to consider the quadratic form $\sigma_{ij}x_ix_j$
(summation over repeated indices) and to show that it is positive for any
nonvanishing $x_i$, $i=1, 2$. And this is the case, since using again the
uncertainty relation (5) we get
\be       %19
\sigma_{ij}x_ix_j \ge \sigma_{11}x_1^2+
\sigma_{22}x_2^2-2|\sigma_{12}x_1x_2| >
(|x_1|\sqrt{\sigma_{11}}-|x_2|\sqrt{\sigma_{22}})^2\ge 0,
\ee
and this ends the proof of Proposition 2.

We note the generality of the above result: the required positivity of
commutator $-i[X,Y]=[A,A^\dg]$ holds for any operators $X, Y$ with
continuous spectrum\cite{D94}, in particular for quadratures of $a^k,
k=1,2,...$ and for generators $K_{1,2}$ of any discrete series reprs of
$SU(1,1)$. In particular case of $X=q, Y=p$ (that is $A = a$) the
Proposition 2 recover the result of refs.\cite{{D95},{Sudarshan}} for
diagonalization of one mode uncertainty matrix by means of linear canonical
transformations. We have to note that the $N$ mode uncertainty matrix
$\sigma(\rho)$ under linear canonical transformations is transformed as in
eq.(18) with $N\times N$ symplectic matrix $\Lambda$\cite{D95} and therefor
can be diagonalized by $\Lambda$ if it is positive definite. In
ref.\cite{D95} it is shown that the $N$ mode $\sigma$ is positive definite
and therefor is diagonalizable in any state. In ref.
\cite{Sudarshan} a diagonalization procedure is described.  It worth
noting another general property of the commutator preserving
transformation (13) (or its equivalent (17)). We can take in it real $u$
and $v$, $u^2-v^2=1$ to get diagonal symplectic $\Lambda(u,v)$ which
performs scaling transformation $X^\pr =
\lambda X, Y^\pr = Y/\lambda$, $\lambda = u+v$. Then when $\lambda
\rightarrow
\infty$ ( $\lambda \rightarrow 0$) we get absolute squeezing of
$Y^\pr $ ($X^\pr$) in any state $\rho$. This in fact stems
from the Proposition 2.

The canonical squeeze operator in quantum optics
$S(\zeta)$\cite{{Caves},{Walls}}, eq.(8) (and its general $SU(1,1)$ form
as well), in fact maps the nonsqueezed $q$-$p$ equal uncertainty HIS
$|z\ra$ (in eq.(6) these are the canonical CS $|\alpha\ra$ ) into $q$-$p$
SIS $|z,u,v\ra$ (in (6) these are the canonical SS $|\zeta,\alpha\ra$).
This observation suggests to define squeeze operator $S(u,v)$ for any
observables $X,\,Y$ as a map HIS $\rightarrow$ SIS,
\be     %20
|z,u,v\ra = S(u,v)|z\ra.
\ee
This is a correct definition of the operator $S(u,v)$ in Hilbert space in
all cases, in which the equal uncertainty HIS $|z\ra$ form an overcomplete
set (a set of general CS\cite{Klauder}) since in those cases any state can
be expressed in terms of $|z\ra$ (for example in cases $A=a$, $A=a^2$,
$A=K_\mp$). In such cases we easily get that the generalized squeeze
operator $S(u,v)$, eq.(20), performs the isometric transformation of
$A^{\pr}$ (proof in the Appendix),
\be      %21
S^{\dg}(u,v)A^{\pr}S(u,v) = A.
\ee
The concepts of squeeze operator is important in practical purposes for
generation of the SIS $|z,u,v\ra$ from HIS $|z\ra$, trying to realize $S$
as a quantum evolution operator to describe the time evolution of initial
$|z\ra$ (when the latter's are available).  In canonical case this method is
effectively applied for generation SS from Glauber CS
$\alpha\ra$\cite{Walls}.  However not always $S(u,v)$ is
unitary, as we shall see below. It is unitary in the canonical case of
$A=a$. When $S(u,v)$ is unitary ($S^\dg S = 1 = SS^\dg$) we have
also $S^\dg A^{\pr2}(u,v)S=A^2$ in addition to eq.(21). Then we get
that the three second moments $\sigma_{ij}(\rho;X^\pr,Y^\pr)$ of
$X^\pr = SXS^\dg$, $Y^\pr = SYS^\dg$ coincide with the
moments of old quadratures $X\,Y$ in the transformed state $\rho^\pr =
S^\dg\rho S$  and similarly the moments of $S^\dg XS,\,S^\dg
YS$ in $\rho$ are equal to those of $X,\,Y$ in $S\rho S^\dg$.
The discussion after the Proposition 2 proves that one can get arbitrary
strong $X$ or $Y$ squeezing  applying $S(u,v)$ to any state $\rho$. Thus
all states $S(u,v)\rho S^{\dg}(u,v)$ are $X$-$Y$ SS. They are SIS if
$\rho$ is SIS or HIS, (HIS $\subset$ SIS), that is if $\rho = |z,u,v\ra\la
v,u,z|$.
\medskip

\bc
{\large {\bf III. Squared amplitude Schr\"{o}dinger and
                  Heisenberg intelligent states }}
\medskip

{\bf A. Equal uncertainty HIS and even and odd  CS}
\ec
In this subsection we provide a brief review of the properties (noting some
new ones) of the even and odd CS (e.\,o.  CS) $|\alpha_{\pm}\ra$
\cite{{DMM},{Perina}} and some other Schr\"odinger cat states in the
light of the Heisenberg uncertainty relation for the quadratures
$X_{sa}, Y_{sa}$ of the squared amplitude $a^2$. The e.\,o. CS, Yurke--Stoler
states\cite{YS} and the recently discussed parity CS\cite{Spiridon} all
are squared amplitude (s.\,a.) equal uncertainty HIS $|z\ra_{sa}$.
Finally we note that in any HIS the three dimensional uncertainty matrix for
the s.\,a.  quadratures $X_{sa},\,Y_{sa}$ and the number operator
$N=a^\dg a$ is diagonal.

The e.\,o. CS $|\alpha_\pm\ra$, which are discussed intensively in the
recent literature as an example of Schr\"{o}dinger cat
states\cite{{Hach},{GGKnight},{Kuang}}, have been first introduced by
Dodonov, Malkin and Man'ko \cite{DMM} as
\be      %22
|\alpha_\pm\ra = N_\pm(|\alpha|)[|\alpha\ra \pm |-\alpha\ra],
\ee
($N_\pm(|\alpha|) = \frac{1}{\sqrt 2}\l(1 \pm \exp\l(-2|\alpha|^2\r)\r)^
{-\frac{1}{2}}$)
where $|\alpha\ra$ is the canonical CS.  These states form overcomplete
families of e.\,o. states,
\be      %23
1_\pm = \frac{1}{\pi}\int d^2\alpha|\alpha_\pm\ra\,\la_\pm\alpha|,
\ee
where $1_\pm$ are the unit operators in the space of e.\,o. states
respectively.  It was noted\cite{DMM} that they are eigenstates of the
squared boson annihilation operator $a^2$, $a^2|\alpha_\pm\ra =
\alpha^2|\alpha_\pm\ra$.  Comparing this equation with eq.(6) we see that
e.\,o. CS are particular case of the SIS $|z,u,v\ra$ with $A=a^2$, $u=1$,
$v=0$ and $z=\alpha^2$. More precisely they are equal uncertainty s.\,a.
HIS (see eq.(4)) for the quadratures $X_{sa},\,Y_{sa}$ of $a^2$,
\be      %24
X_{sa}=\frac{1}{\sqrt2}(a^2+a^{\dg 2}),\quad Y_{sa} =
-\frac{i}{\sqrt2}(a^2-a^{\dg 2}),
\ee
\be      %25
[A,A^\dg]=-i[X_{sa},Y_{sa}] = [a^2,a^{\dg^2}] = 4a^\dg a + 2,
\ee
The mean number operator and the variances are
\be      %26
\la_{+}\alpha|a^\dg a|\alpha_{+}\ra = |\alpha|^2{\rm
tanh}\,|\alpha|^2, \quad
\la_{-}\alpha|a^\dg a|\alpha_{-}\ra = |\alpha|^2{\rm
ctanh}\,|\alpha|^2,
\ee
\be      %27
\Delta^2_{X_{sa}}(\alpha) = \Delta^2_{Y_{sa}}(\alpha) = 1 + 2\la a^\dg
a\ra,\quad \Delta_{X_{sa}Y_{sa}} \equiv 0.
\ee
In both type of states the lowest level $\Delta_0$ of the equality (4) for
the operators $X_{sa},\,Y_{sa}$ is $\Delta_0=1$ and is reached at $\alpha
= 0 = z$. We see that both variances $\Delta_{X_{sa}}(\alpha)$ and
$\Delta_{Y_{sa}}(\alpha)$ are greater than $\Delta_0=1$.  Therefor the
e.\,o. CS are not amplitude-squared SS neither according to the definition
(3), nor according to the relative definition (2).  It worth to note that
due to the positivity of the commutator $[a^2,a^{\dg 2}]$ any s.\,a.
equal HIS  is an eigenstate of $a^2$.  Then due to the linearity of the
eigenvalue equation $a^2|z\ra_{sa} = z|z\ra_{sa}$ ( see eq.  (40) below) the
general form of the s.\,a. equal uncertainty HIS is a linear
combination of one even and one odd CS,
\be      %28
|z\ra_{sa} = C_+(\alpha)|\alpha\ra_+ + C_-(\alpha)|\alpha\ra_-,
\ee
where $|C_-(\alpha)|^2 + |C_+(\alpha)|^2 =1$ and the eigenvalue
$z=\alpha^2$. Important physical examples of equal uncertainty HIS (28)
are the Yurke-Stoler states \cite{YS}
\be      %29
|\alpha;YS\ra = \frac{1}{\sqrt 2}\l(\exp\l(-i\pi/4\r)|\alpha\ra
+ \exp\l(i\pi/4\r)|-\alpha\ra\r),
\ee
and more general Spiridonov parity CS $|\alpha\ra_p$ \cite{Spiridon},
defined as eigenstates of the product $Pa$, where $P$ is the operator of
inversion (parity operator) and $a$ is boson destruction operator. Indeed,
let $Pa|\alpha\ra_p = \alpha|\alpha\ra_p$.  Then from $PaP = -a$, $P^2 =
1$ we get $(Pa)^2 = PaPa = -a^2$ and therefor $|\alpha\ra_p$ is eigenstate
of $a^2$.  The inverse is not true, i.e. parity CS $|\alpha\ra_{p}$ are
particular case of s.\,a. HIS $|z\ra_{sa}$.

The nonclassical properties of e.\,o. CS $|\alpha\ra_\pm$ have been
considered for example in refs.
\cite{{BuzekJeh},{Vinogradov},{Barranco}}. Note that in e.\,o. CS
the second moments of $q$ and $p$
do not yield the equality in the Schr\"{o}dinger relation (5), i.e. they
are not $q$-$p$ SIS.

Consider briefly the uncertainty matrices in HIS $|z\ra_{sa}$.  The
$q$-$p$ uncertainty matrix in s.\,a. HIS can be or can not be diagonal.
For example in e.\,o. CS the covariance is $\Delta_{qp}(\alpha_\pm) = {\rm
Im}\alpha^2$ in both $|\alpha_\pm\ra$. Contrary to this the
$X_{sa}$-$Y_{sa}$ uncertainty matrix (and even the $3\times3$ s.\,a.
uncertainty matrix for the three operators $X_{sa},\,Y_{sa},\,a^\dg
a$) is diagonal in all HIS $|z\ra_{sa}$ and this can be checked directly,
using the eigenvalue property (15).  The nonvanishing diagonal elements
are
\be      %30
\Delta^2_{X}(z)=\Delta^2_{Y}(z)=\frac{1}{4}(2\la a^\dg a\ra+1);\quad
\Delta^2_{a^\dg a}(z) = |z|^2 + \la a^\dg a\ra^2.
\ee
The Proposition 2 in the preceding section guaranties that the $2\times2$
s.\,a. uncertainty matrix $\sigma(\rho;X_{sa},Y_{sa})$ can always be
diagonalized by commutator preserving linear transformation (13). For
$\sigma(\rho;X_{sa},Y_{sa})$ there is one more possibility to be
diagonalized - this is by means of linear canonical transformations (note
that the latter's are not s.\,a. commutator preserving ones).

{\it Proposition 3}. The $2\times2$ squared amplitude uncertainty matrix in
any pure or mixed state $\rho$ ca be diagonalized by linear canonical
transformation.

Proof. Let us consider three independent quadratic boson operators (the
standard $SU(1,1)$ generators) $K_i$, $i=1,2,3$ ,
\be      %31
K_1 = \frac{1}{4}(a^2 +a^{\dg 2}),\,\,K_2 = i\frac{1}{4}(a^2
-a^{\dg 2}), \,\, K_3 = \frac{1}{4}(2a^{\dg}a+1).
\ee
Linear canonical transformations
\be     %32
a \rightarrow a^\pr = \mu a + \nu a^\dg,\quad |\mu|^2-|\nu|^2=1
\ee
are generated by the unitary methaplectic operators (summation over
repeated indices)
\be      %33
U(\vec{\zeta}) = \exp\l( i\zeta_j\,K_j\r),\quad
\vec{\zeta}=(\zeta_1,\zeta_2,\zeta_3),
\ee
where $\zeta_i$ are $3$ real parameters.
This operators act on the components $K_i$ as (pseudo) rotations in
Minkovski space $M_3$ and they contain rotations in the space like
plain of $K_1$ and $K_2$,
\be      %34
U^{\dg}(\vec{\zeta})K_jU(\vec{\zeta}) \equiv K^\pr_j =
\lambda_{jl}(\vec{\zeta})K_l,
\ee
where the Lorentz matrices $\Lambda(\vec{\zeta}) = (\lambda_{jl})$  obey
the relation $\Lambda^{T}(\vec{\zeta})g\Lambda(\vec{\zeta})=g$, $g$ being
the metric tensor. Under such transformation the $3\times3$ matrix
$\sigma(\rho;K_1,K_2,K_3)$ transforms as a second rank symmetric tensor
($\sigma^\pr = \sigma(\rho;K_1^\pr,K_2^\pr,K_3^\pr$),
\be      %35
\sigma^\pr(\rho)
=\Lambda(\vec{\zeta})\sigma(\rho)\Lambda^{T}(\vec{\zeta}).
\ee
The matrix $\sigma(\rho;K_1,K_2)$ is the left upper $2\times2$ block and
under rotations in the space like plane transforms according to (18) with
orthogonal matrix $\Lambda$. Therefore it can always be brought to diagonal
form by means of linear canonical transformations. End of proof.

Thus linear canonical transformations can diagonalize both amplitude
(canonical) and squared amplitude $2\times2$ uncertainty matrices. Let us
point out states which are not SIS but in which the s.\,a. matrix $\sigma$
is diagonal: those are Fock states $|n\ra$. In the latter states all
covariances $\Delta_{K_iK_{k\neq i}}$ are vanishing and the variances are
\be      %36
\Delta^2_{K_1}=\Delta^2_{K_2}=\frac{1}{2}(n^2+2n+\frac{1}{2}),\quad\,
\Delta^2_{K_3}=0.
\ee
The diagonalization of uncertainty matrix $\sigma$ of any $n$ observables
$X_i$ is a minimization of the Robertson inequality\cite{Robertson34}
$\Delta^2X_1\Delta^2X_2...\Delta^2X_n \geq  {\rm det}\sigma$.
\medskip
\bc
{\bf B. Squared amplitude SIS and generalized even and odd CS }
\ec
\medskip
In the previous subsection we have written down explicitly the general
form of the s.\,a. equal uncertainty HIS $|z\ra_{sa}$ as superposition of
one even and one odd CS (eq. (20)).  Our aim now is to find general
solution for the s.\,a. SIS $|z,u,v\ra_{sa}$ in quite analogous form. We
shall construct two independent sets of even and odd SIS which could be
considered as a generalization of e.\,o.  CS $|\alpha_{\pm}\ra$. From the
point of view of boson squeezing the aim is to obtain nonclassical states
that can exhibit strong amplitude-squared squeezing (quadratic squeezing)
in the sense of definition of eq.(3) and ordinary squeezing (linear
squeezing or $q,p$ squeezing) as well.

According to the discussion in the preceding section we are looking for
normalizable solutions $|z,u,v\ra$ of the eigenvalue problem (6) with
$A=a^2$.  The commutator $[A,A^\dg]$ now is given by the eq.(25),
which shows that it is positive definite. Then the normalized solutions of
eq.(6) could exist for $|v/u| > 1$ only \cite{D94}.
Thus we have to solve the eigenvalue equation
\be      %37
(ua^2+va^{\dg 2})|z,u,v\ra_{sa} = z|z,u,v\ra_{sa}.
\ee
We shall solve this equation using the canonical CS representation
\cite{Klauder}. In this representation pure states $|\psi\ra$ are
represented by entire analytic functions of order $1/2$ and type $2$
(quadratic exponent type), i.e. of growth $(1/2,2)$,
\be      %38
|\psi\ra
\longrightarrow f_\psi(\alpha^*) = {\rm exp}(-|\alpha|^2/2)\la\alpha|\psi\ra,
\ee
and the boson destruction and creation operators are
\be      %39
a = \frac{d}{d\alpha^{*}},\quad a^\dg = \alpha^{*}.
\ee
The number states $|n\ra$ are represented by $(\alpha^*)^n/\sqrt{n!}$ and
the Glauber CS $|\beta\ra$ by ${\rm exp}\l(-|\beta |^2 + \beta\alpha^*\r)$.
So  in  canonical CS representation the eigenvalue eq.(37) is the following
second order differential  equation,
\be      %40
(u\frac{d^2}{d{\alpha^*}^2} + v{\alpha^*}^2 - z)\Phi(\alpha^*) = 0.
\ee
This equation is easily reduced to the Kummer hypergeometric equation
\cite{Stegun}.  Then we have the following two independent solutions of
eq.(40)
\be      %41
\Phi _{+}(\alpha ^{*};z,u,v) = {\cal N}_{+}\,{\rm
exp}\left(-\frac{1}{2} {\tilde{\alpha ^{*}}}^{2}\right)\, _{1}F_{1}\left(
a_{+}, \frac{1}{2}; {\tilde{\alpha ^{*}}}^{2}\right)
\equiv {\cal N}_{+}\tilde{\Phi} _{+}(\alpha ^{*}),
\ee
\be      %42
\Phi _{-}(\alpha ^{*};z,u,v) = \alpha ^{*}
{\cal N}_{-}\,{\rm exp}\l(\frac {1}{2}{\tilde{\alpha ^{*}}}^{2}\r)\,
_{1}F_{1}\l(a_{-},{3\over 2};-{\tilde{\alpha^{*}}}^{2}\r)
\equiv {\cal N}_{-}\tilde{\Phi} _{-}\alpha ^{*}),
\ee
where ${\cal N}_{\pm}$ are normalization constants (they are functions of
$z,u,v$),
$$\tilde{\alpha ^{*}} = \alpha^* (-v/u)^{1/4},\quad a_{+} = \frac{1}{4}(
1 + z/\sqrt{-uv}),\quad a_{-} = \frac{1}{4}(3 + z/\sqrt{-uv}), $$
and
$_{1}F_{1}(a,b;z)$ is the Kummer confluent hypergeometric function
\cite{Stegun}. For $b\neq -n$ (as in our case, where $b=1/2$ or $b=3/2$)
it is an entire analytical function of $z$ which for $|z| \rightarrow
\infty$ increases not faster than exp$(|z|)$.  Thus the properties  of
$_{1}F_{1}(a,b=1/2,3/2;z)$ ensure the required  growth  and  analyticity
of solutions (28),\,(29) so that they represent normalizable states
$|z,u,v;\pm\ra$ provided the inequality $|v/u|\, < \, 1$ holds. This
is an explicit example of the general statement that if the commutator
$[A,A^\dg]$ is positive, then the normalizable SIS exist for
$|v/u|\,<\,1$ only\cite{D94}. The explicit realizations of the above IS
$|z,u,v;\pm\ra$, eqs. (41,42), demonstrates, that up to a phase factor
they are determined by the two complex parameters $z/u$ and $v/u$, so that
we can write  $|z,u,v;\pm\ra \, =
\,|z/u,v/u;\pm\ra$. The normalization constants also depend (up to phase
factors) on $z/u,\,v/u$ only and are given by the integrals
\be      %43
{\cal N}_{\pm }^{-2}(z,u,v) = \frac{1}{\pi}\int d^2\alpha
\tilde{\Phi}_{\pm}^*(\alpha ^{*};z,u,v)\tilde{\Phi}_{\pm}
(\alpha ^{*};z,u,v) \equiv I_{N_{\pm}},
\ee
which are convergent due to the right analytical properties of solutions
$\Phi_\pm(\alpha^*)$. Solutions (41) and (42) were obtained and
briefly discussed in \cite{D95}.

The above solutions $\Phi_\pm(\alpha^*)$ are e.\,o. functions of
$\alpha^{*}$ respectively so that they represent the e.\,o. SIS. The
odd states $|z,u,v;-\ra$ are orthogonal to the even states $|z,u,v;+\ra$.
Since the eigenvalue eq.(37) is necessary and sufficient condition for a
state to be s.\,a. SIS and since it is a linear second order
differential equation we have the result  that {\it any} s.\,a.
SIS $|z,u,v\ra_{sa}$ is a solutions of the eq.(37) and takes the form of
linear combination of one even and one odd SIS $|z,u,v;\pm\ra$,
\be      %44
|z,u,v\ra_{sa} = {\cal C}_-(z,u,v)|z,u,v;-\ra + {\cal C}_+(z,u,v)|z,u,v;+\ra,
\ee
where $|{\cal C}_-|^2 + |{\cal C}_+|^2 = 1$. This generalizes the HIS
relation (28) to the case of SIS.

Let us consider some interesting subsets of the  e.\,o. SIS $|z,u,v;\pm\ra$.
First of all, as eq.(37) shows,  when $v=0$  the states $|z,u=1,v=0\ra_{sa}$
are eigenstates of $A = a^2$, that is $|z,u=1,v=0;\pm\ra$ have to coincide
with some subset of s.\,a. equal uncertainty HIS $|z\ra_{sa}$. The precise
subset of equal uncertainty HIS is obtained by substitution $u=1,\,v=0$ in
solutions (41) and (42). We have
$$
\tilde{\Phi}_{+}(\alpha ^{*};z,1,0) = {\rm cosh}(\alpha^*\sqrt z),\quad
\tilde{\Phi}_{-}(\alpha ^{*};z,1,0) = {\rm sinh}(\alpha^*\sqrt z),
$$
which proves that $|z,1,0;\pm\ra$ coincide with the Dodonov et. all
e.\,o.  CS\cite{DMM} $|\beta_\pm\ra$ with $\beta = \sqrt z$ since
functions sinh$(\beta\alpha^*)$ and cosh$(\beta\alpha^*)$ represent (up to
normalization factors) the e.\,o. CS $|\beta_\pm\ra$ in the canonical CS
representation respectively.  The ordinary e.\,o. CS form overcomplete
family of states in the sense of the resolution of unity operator,
eqs.(23), (overcompleteness in the strong sense \cite{Klauder}).  Then the
SIS $|z,u,v;\pm\ra$ form at least dense set in Hilbert space and according
to ref.\cite{Klauder} could be called CS (at least) in a weak sense. This
is the motivation to call $|z,u,v;\pm\ra$ {\it generalized e.\,o.} CS.
Similarly any specific combination $C_-|z,u,v,-\ra + C_+|z,u,v,+\ra$ could
be considered as a generalization of the corresponding amplitude-squared
equal uncertainty HIS (28), in particular of Yurke-Stoler states or any
parity CS $|\alpha\ra_p$.

We note that the s.\,a. squeeze operator $S_{sa}(u,v)$, defined according
to general definition (20) is isometric, but not unitary, (proof in the
Appendix) and therefore the sets of SIS $|z,u,v;\pm\ra$ do not resolve the
unity operators $1_-$ and $1_+$ (eq.(23)) by integration of the projectors
$|z,u,v;\pm\ra\la\pm,v,z|$ against the old measure
$d\mu(\alpha)=(1/\pi)d^2\alpha$, $z=\alpha^2$: the integration against
this measure yields orthogonal projectors $P_\pm(u,v)$ on the linear span
of SIS $|z,u,v;\pm\ra$ with fixed $u,\,v$,
\be      %45
P_\pm(u,v) = \frac{1}{\pi}\int
d^2\alpha|\alpha^2,u,v;\pm\ra\la\pm;v,u,\alpha^2|.
\ee

The second subsets we note here are rather unexpected: those are the sets
of canonical squeezed vacuum states $|\zeta;0\ra$ (in Perelomov notation
$|\xi;0\ra$, $|\xi|^2 < 1$) and squeezed one photon states $|\xi;1\ra$,
\be      %46
|\xi;0\ra = (1-|\xi|^2)^{\frac{1}{4}}\exp\l(\xi a^{\dg 2}\r)|0\ra,\quad
|\xi;1\ra = (1-|\xi|^2)^{\frac{3}{4}}\exp\l(\xi a^{\dg 2}\r)|1\ra.
\ee
These states are recovered by our generalized e.\,o. CS $|z,u,v;\pm\ra$ when
the following relations between parameters $z,\,u$ and $v$ are imposed
\be      %47
z=\pm\sqrt{-uv} \qquad {\rm or}\qquad z = \pm 3\sqrt{-uv}.
\ee
Substituting these into solutions (41), (42) we get ($w\equiv v/u$)
\be      %48
\tilde{\Phi}_{-}(\alpha^{*}) = \alpha^*\exp\l(\mp{\alpha^*}^2\sqrt{-w}\r),
\quad  \tilde{\Phi}_{+}(\alpha^{*}) = \exp\l(\mp{\alpha^*}^2\sqrt{-w}\r),
\ee
which coincide (up to normalization factors) with the Glauber CS
representation of $|\xi;0\ra$ and $|\xi;1\ra$ respectively with $\xi =
\pm\sqrt{-w}$. This proves that the Perelomov $SU(1,1)$ CS with Bargman
index $k=1/4,\,3/4$ minimize the Schr\"odinger u.\,r. for the generators
$K_{1,2}$, For the cases of any square integrable representation ${\cal
D}^\pm(k),\quad k=1/2,1,..., $ the above property of $S(1,1)$ CS was
established in\cite{D94} using the representation of Barut--Girardello
CS\cite{BG}. It worth noting that if in the Barut-Girardello
representation $\la k;z^{\pr}|z,\lambda;k\ra$ of $SU(1,1)$ SIS
$|z,\lambda;k\ra$, constructed in ref.\cite{D94}, we put $k=1/4$,
$z^{\pr}=\alpha^2/2$ we would get the s.\,a. SIS (41,42) in Glauber CS
representation in spite of the fact that the Barut-Girardello representation
is correct for Bargman indices $k=1/2,1,...$ only.

Let us recall the known fact that the  $SU(1,1)$ CS $|\xi;0\ra$ (equal to
canonically squeezed vacuum $|\zeta;0\ra$) are also eigenstates of the
linear combination $\mu a + \nu a^\dg$, $|\mu|^2 - |\nu|^2 =1$, and
therefor satisfy the equality in Schr\"odinger u.\,r. for the amplitude
quadratures $q,\,p$.  Thus $SU(1,1)$ CS $|\xi;0\ra$ are very symmetric
family of boson field states which minimize the Schr\"odinger u.\,r. for
both amplitude and s.\,a.  quadratures (double SIS). It is known that they
exhibit strong linear amplitude squeezing according to the definition (3).
But they do not exhibit amplitude-squared squeezing: one can check that
the variances $\Delta X(\xi)$ and $\Delta Y(\xi)$ for any $\xi \ne 0$ are
greater than $\Delta_0 = 1$. Only relative squeezing in the sense of
eq.(2) holds: the ratio $2(\Delta X)^2/|\la[X,Y]\ra|$ here is equal to
$|1+\xi^2|^2/(1+|\xi|^2)^2$ and when $\xi \rightarrow \pm i$ it tends to
$0$ (i.e. 100\% relative squeezing).

A third type of  particular cases of SIS $|z,u,v;\pm\ra$ is  obtained
when $z$ is related to $u$, and $v$ according to
\be      %49
z = -(4n+1)\sqrt{-uv} \quad {\rm or}\quad z= -(4n+3)\sqrt{-uv} ,
\ee
\ni where $n$ is positive integer. In this cases the Kummer function
$_{1}F_{1}(a,1/2;z^{2}/2)$ to within a constant factor coincides with
Hermite polynomial $H_{2n}(z)$ and $_{1}F_{1}(a,3/2;z^{2}/2)$ to within a
factor coincides  with $(1/z)H_{2n+1}(z)$ \cite{Stegun}. Then from the
explicit form of solution (41), (42) and the representation (39) we derive
that the e.\,o. SIS $|z,u,v;\pm\ra$ under the restrictions (49) take the
form of {\it finite superposition} of ordinary squeezed number states
$|n\ra$.  If  furthermore in addition to (49) we take in $|z,u,v;\pm\ra$
$u$ and $v$ real,
\be      %50
u = \frac{1}{2}(1+\lambda),\quad v = \frac{1}{2}(1-\lambda),\quad \lambda
>0,
\ee
we  would get the "minimum uncertainty states for amplitude-squared
squeezing" (or the squeezed Hermite polynomial states) considered in
refs.\cite{BHYu}. The latter's constitute a subset of s.\,a.  HIS.  Our
squeezed Hermite polynomial states, corresponding to restrictions (49) in
$|z,u,v;\pm\ra$ only, (being SIS) are more general - they e.g. admit
nonvanishing $X_{sa}$-$Y_{sa}$ covariance (correlated\cite{DKM} Hermite
polynomial states).

Using the correspondence (39) we can write the even and odd s.\,a. SIS as
double series in terms of Fock states,
\be      %51
|z,u,v;+\ra = {\cal
N}_{+}\sum_{n=0}^{\infty}g_{2n}|2n\ra,\quad
|z,u,v;-\ra = {\cal N}_{-}\sum_{n=0}^{\infty}g_{2n+1}|2n+1\ra,
\ee
where ($(a)_{k}$ is the Pohgammer symbol)
\bear*
g_{2n}=\sqrt{(2n)!}\sum_{k=0}^{n}\l(-\frac{1}{2}\sqrt{-\frac{v}{u}}\r)^{n-k}
\frac{\l(a_{+}\r)_{k}}{\l(\frac{1}{2}\r)_{k}},\\
g_{2n+1} =
\sqrt{(2n+1)!}\sum_{k=0}^{n}\l(-\frac{1}{2}\sqrt{-\frac{v}{u}}\r)^{n-k}
\frac{\l(a_{-}\r)_{k}}{\l(\frac{3}{2}\r)_{k}}.
\eear*
In conclusion to this section it worth noting that the functions (41), (42)
are solutions of the eigenvalue equation (40) for any complex parameters
$u, v$ and $z$. When $|v/u|<1$ they represent normalized states. When
$|v/u|\geq1$ these solutions could be considered as nonnormalizable states.
For example when $u, v$ are real and $v=\pm u$ we get the eigenfunctions of
continuous observables $X_{sa}$ and $Y_{sa}$. The situation is the same in the
canonical case of $q$ and $p$: the canonical SS are solutions of eigenvalue
equation (6) with $A=a$ and if we write down this equation in CS
representation we would get nonnormalized solutions for $u=\pm v$ which
could be considered as nonnormalized eigenstates of $q$ and $p$ (recall the
nonnormalizable plain wave as eigenstate of $p$ in $q$ representation).

\bc
\bigskip
{\large{\bf IV. Squeezing and photon distributions in squared amplitude IS}}
\ec
\medskip
The constructed e.\,o. squared amplitude SIS $|z,u,v;\pm\ra$ turned out to
exhibit strong both linear (ordinary) and quadratic amplitude squeezing and
to show super- and subpoissonian photon statistics as well.
Here we consider these properties explicitly.

The possibility for joint linear and quadratic amplitude squeezing stems
from the spectral properties of the commutators between quadratures of $a$
and of $a^2$. Indeed, let ${\tilde q}$ denotes $q$ or $p$ and ${\tilde
X}$ denotes $X$ or $Y$. Then consider the commutator
$[{\tilde q},{\tilde X}]$. It is proportional to ${\tilde q}$ and
therefor is non positive and non negative definite. Then states
 $|\psi\ra$ exist in which the mean of this commutator vanishes. Such are
all e.\,o. states for example. Thus for the quadratures ${\tilde q}$
and ${\tilde X}$ Heisenberg relation for such states reads
\be      %52
\Delta^2_{\tilde q}\,\Delta^2_{\tilde X} \ge 0,
\ee
which means that in these states both variances $\Delta{\tilde q}$ and
$\Delta {\tilde X}$ could be simultaneously small (but only one could tend
to zero). There is no restriction from the above as well, i.e. these
variances could be simultaneously large. A similar inequality holds for the
variances of $\Delta{\tilde q}$  and the number operator $N=a^\dg a$,
\be      %53
\Delta^2_{\tilde q}\,\Delta^2_{N} \ge 0,
\ee
which explains the nonexistence of any relation between super- or
subpoissonian statistics and $q$ or $p$ squeezing\cite{Walls}.

\bc
\bigskip

{\bf A. Linear squeezing in squared amplitude SIS }
\ec
\medskip

The variances of the canonical operators $q,\,p$ and their covariance in
any s.\,a. SIS $|z,u,v\ra$ can be easily obtained in terms of the means of
$a^\dg a$ and $q$ and $p$ (expressing $q,\,p$ in terms of
$A^\pr(u,v),\,A^{\pr\dg}$ and taking into account the eigenvalue
eq.(37)). In the e.\,o. states $|z,u,v;\pm\ra$ these formula simplify due
to the orthogonality $\la +,v,z|z,u,v,- \ra = 0$ which lead to $\la q\ra =
0 = \la p\ra$.  %
\be      %54
\Delta^2_q(z,u,v) = \la q^2\ra = \frac{1}{2} + \la v,u,z|a^\dg
a|z,u,v\ra + {\rm Re}[(u-v)z^*],
\ee \be      %55
\Delta^2_p(z,u,v) = \la p^2\ra = \frac{1}{2} + \la v,u,z|a^\dg
a|z,u,v\ra - {\rm Re}[(u-v)z^*],
\ee \be      %56
\Delta_{qp}(z,u,v) = \frac{1}{2}\la qp+pq\ra = {\rm Im}[(u-v)z^*].
\ee
The means $\la\pm,v,z|a^\dg a|z,u,v;\pm\ra$ can be calculated in the
canonical CS representation, using (39), (41) and (42). Since the
integrals $I_{N_{\pm}}(z,u,v)$ cannot be expressed in a simple closed form
in terms of known special functions we have to resort to numerical
calculations or to analytical approximation. Using the known formula
$$
\int \exp\l(-|\alpha|^2\r)\alpha^n \alpha^{*m}d^2\alpha =
\pi n!\delta_{n,m} $$
we can convert the integrals into series and regroup the terms in an
appropriate way to obtain good approximations. For example in the case of
even states $|z,u,v,+\ra$ we get
\be      %57
\la+,v,z|a^\dg a|z,u,v,+\ra =
[I_{N_{+}}(z,u,v)]^{-1}\,\sum_{n=0}^{\infty}2n(2n)!|f_n(z,u,v)|^2|v/u|^n\,,
\ee  \be      %58
I_{N_{+}}(z,u,v)] = \sum_{n=0}^{\infty}(2n)!|f_n(z,u,v)|^2|v/u|^n\,,
\ee
where
$$f_n(z,u,v) =
\sum_{k=0}^{n}(-\frac{1}{2})^k\frac{(a_+(z,u,v))_{n-k}}{k!(n-k)!
(\frac{1}{2})_{n-k}}, \quad a_+(z,u,v) = \frac{1}{4}(1+z/\sqrt{-uv})\,,
$$
$(a)_{n}$ being the Pohgammer symbol.  Both series (57) and (58) are with
positive terms and are convergent by construction. Therefor their first
several terms can be used as approximations to the exact functions for
small $z/u$ and $v/u$ (recall that the normalized eigenstates $|z,u,v\ra$
exist for $|v/u|<1$ only).

In this manner we find that in the family of e.\,o. SIS $|z,u,v;\pm\ra$
there are states which exhibit very strong linear amplitude squeezing. As
illustration we show on Fig.1 the plots of the variance of $q$ in the
three even SIS $|z,u,v;+\ra$ with
$z=-1,-2,-5,\, v=-x,\quad x>0 $
as functions of $x$, where $x$ is positive.  We see that the variance of
$q$ has broad and well pronounced minimums which are deeper for large
(real) $z$. This $q$--squeezing is very strong (much stronger than in
ordinary e.\,o. CS $|\alpha_\pm\ra$) -- the squared variance
$\Delta^2_q(z,u,v)$ for $z=-5,\, v=-5,\, u=\sqrt{26}$ is less than
$0.0025$, and the ratio $\Delta^2_q(z,u,v)/\Delta^2_0$  -- less than
$0.01$ (99\% squeezing). Similar to this (but not identical) is the
squeezing of $p$ in the symmetric states $|z,u,v;\pm\ra$ with
$z=1,2,5,\quad\, v=-x,\quad x>0.$
The photon statistics in these particular $q$-$p$ squeezed even SIS is
superpoissonian (positive Mandel $Q$-factor). On Fig.3 we show the photon
distribution $f(n)$ in the strongly $q$-squeezed state
$|-5,\sqrt{37},-6;+\ra$.  In the large family of even SIS $|z,u,v;+\ra$
there are states with subpoissonian photon statistics ($Q<0$). Those are
for example SIS $|z=\pm5,u=\sqrt{1+|v|^2},v= x;+\ra$ when
$0\,<\,x\,<\,0.5$ (on Fig.3 we show $f(n)$ in $|-5,\sqrt{1.04},0.2\ra$).
Recall that for $v=0$ in the ordinary even CS $|\alpha_+\ra$ the
statistics is superpoissonian for any $\alpha \ne 0$.  It is subpoissonian
in the odd CS $|\alpha_-\ra$, which however are not $q$-$p$ squeezed. Thus
the e.\,o. SIS exhibit more $q$-$p$ squeeze and statistical properties.
Squeezing in some SIS from the subset, determined by the restrictions (49)
and (50) have been considered in ref.\cite{BHYu} (but in the sense of
definition (2), not (3)). The Mandel $Q$-factor,
$Q = \la a^\dg a\,a^\dg a\ra/\la a^\dg a\ra - \la a^\dg a\ra - 1,$
in any SIS $|z,u,v\ra_{sa}$ is given by the formula
\be      %59
Q(z,u,v) = 4|v|^2 - \la a^\dg a\ra + \frac{1}{\la a^\dg
a\ra}\l(2|v|^2+|z|^2(1+2|v|^2)-2\,{\rm Re}(u^*v^*z^2)\r).
\ee
We see that subpoissonian statistics is most likely to occur in SIS with
${\rm Re}(u^*v^*z^2) > 0$.

\bc
\bigskip
{\bf B. Quadratic squeezing in squared amplitude SIS }
\ec
\medskip
The three second moments of the s.\,a. quadratures in our SIS
$|z,u,v\ra_{sa}$ are obtained from the general $X$-$Y$ formula,
eqs.(9-11), in the form
\be      %60
\Delta^{2}_{X_{sa}}(z,u,v)\,=\,|u-v|^2(2\la a^\dg a\ra + 1),
\ee    \be      %61
\Delta^{2}_{Y_{sa}}\,=\,|u+v|^2\l( 2\la a^\dg a + 1 \ra\r),\quad
\Delta_{XY}\, = \,4{\rm Im}(u^*v)\l(2\la a^\dg a\ra +1\r).
\ee
From these formula and from the eigenvalue equation for s.\,a. SIS it
follows that in the limits $v\rightarrow \pm u$ (possible only when $|v|
\rightarrow \infty$) the variances of $X_{sa}$ or $Y_{sa}$ should become
arbitrary small. For finite $v$ we get finite quadratic squeezing.
%%%\bc \input{fig1-2} \ec \vspace{0.3cm}
On Fig. 2
we show the plots of the variance of $Y_{sa}$ as function of $x$ in the
strongly $q$ squeezed even SIS $|z,\sqrt{1+x^2},-x;+\ra$, $z=-1,-2,-5$ and
$x>0$.  The quadratic squeezed vacuum states $|0,\sqrt{1+x^2},-x;+\ra$
(which are annihilated by $ua^2+va^{\dg 2}$ and at $v=0$ coincide with
the true vacuum $|0\ra$) also exhibit strong s.\,a. squeezing. Here we have
to point out that e.g. the states $|-5,\sqrt{1+x^2},-x;+\ra$ in the
interval $4.5< x <8$ are $q$- and $Y_{sa}$--squeezed
simultaneously (joint $q$ and $Y_{sa}$ squeezing).  Symmetric to these are
states $|5,\sqrt{1+x^2},-x;+\ra$ which are also double squeezed, this time
$p$ and $X_{sa}$ being squeezed simultaneously.  One can expect that the
squeezed states with joint linear and quadratic squeezing should be useful
in optical communications to reduce further the noise of the field.

For any quantum state  we can establish simple geometric condition, which
is sufficient for the state to exhibit superpoissonian photon statistics.
This condition is shown to be more efficient for s.\,a. equal uncertainty
states and necessary and sufficient for s.\,a. equal uncertainty HIS
$|z\ra_{sa}$, eq.(28). The possibility for such relation is provided by the
Casimir operator of $SU(1,1)$ for the representation (31): $C \equiv K_1^2
+ K_2^2 - K_3^2 = -k(k-1) = 3/16$.  Using this we can express the variance
of number operator $N=a^\dg a = 2K_3 - 1/2$ in terms of the variances
of $X_{sa}$, $Y_{sa}$ and the components of the mean quasi spin vector
$\la\underline{K}\ra\equiv (\la K_1\ra,\la K_2\ra,\la K_3\ra)$,
\be      %62
\la\underline {K}\ra^2 = \la K_{1}\ra^2 + \la K_{2}\ra^2 - \la K_{3}\ra^2,
\ee
and thus to establish relation between the Mandel Q-factor and the squared
amplitude variances. We have (in any state)
\be      %63
\la N\ra Q =
4\l(\Delta^2_{K_1}+\Delta^2_{K_2}\r)+4\la\underline{K}\ra^2-2\la K_3\ra
-\frac{1}{4}.
\ee
%%% \bc \input{fig3-4} \ec \vspace{0.2cm}
Using the Schr\"odinger relation (5) we get the desired sufficient condition
\be      %64
\la\underline{K}\ra^2 \, \ge \, \frac{1}{16} - \frac{1}{2}\la K_3\ra.
\ee
This condition involves first moments of $K_i$ only and has the further
advantage that the "length" of mean quasi spin vector
$\la\underline{K}\ra^2$ is invariant under linear canonical
transformations,
\be      %65
\la \psi,\vec{\zeta}|\underline{K}|\vec{\zeta},\psi\ra^2 \,=\,\la
\psi|\underline{K}|\psi\ra^2,
\ee
\be      %66
|\vec{\zeta},\psi\ra \,=\,U(\vec{\zeta})|\psi\ra,
\ee
$U(\vec{\zeta})$ being the methaplectic operator (33), which generates
linear canonical (homogeneous for simplicity) transformations and when
$\zeta_3 = 0$ coincides with the canonical squeeze operator $S(\zeta)$,
eq.(8). Thus in all states of the form $U(\vec{\zeta})|\psi\ra$ we have to
calculate in fact the mean of $K_3$ and the mean quasi spin in $|\psi\ra$
only. Condition (64) is satisfied by squeezed even CS
$S(\zeta)|\alpha_+\ra$ when the quantity $\sinh r\,{\rm
Re}(\alpha^2e^{i\theta})\ge 0$ ($\zeta = re^{i\theta}$), in particular by
all even CS $|\alpha_+\ra$. Squeezed number states $|\zeta,n\ra$ satisfy
the above condition when $\sinh^2r + \cosh^2r \ge n+1$.

It worth noting that neither ordinary squeezed CS $|\zeta,\alpha\ra$ nor
ordinary squeezed number states $|\zeta,n\ra$ exhibit quadratic squeezing
in the sense of eq.(3) what could be verified  after some tedious
calculations.  Only quadratic squeezing after definition (2) could
exist in $|\zeta,\alpha\ra$ and $|\zeta,n\ra$. So the SIS
constructed here are probably the first examples of squeezed states
with joint amplitude and amplitude-squared squeezing.

\bigskip
\bc
{\large{\bf V. On the stable evolution and generation of SIS }}
\ec
\medskip
In this section we consider some aspects of the problem of time evolution
of initial SIS $|z_0,u_0,v_0\ra$ in grater detail treating the s.\,a. SIS.
We discuss possible generation of s.\,a. SIS and SS from other known
states.

If $U(t)$ is an evolution operator for a given
quantum system then the time evolution of the initial SIS is
$U(t)|z_0,u_0,v_0\ra = |t;z_0,u_0,v_0\ra$. The evolution is called stable
if $|t;z_0,u_0,v_0\ra$ is again SIS. That is (up to a phase factor)
\be      %67
|t;z_0,u_0,v_0\ra = |z(t),u(t),v(t)\ra,
\ee
where complex parameters $z(t)\equiv z,\, u(t)\equiv u,\, v(t)\equiv v$ are
functions of time. Physical importance of stable evolution
of a given set of states is in that such states can be realized for the
system described by $U(t)$ and they can be generated by acting with $U(t)$
on some known states from the same set. If the evolution is stable for a
subset only, then other states from the set can not be realized for this
system (the time evolution operator would destroy such states).

For a quantum system with Hamiltonian $H$ (possibly time dependent) the
evolution of a given set of SIS $|z,u,v\ra$ is stable if the following
(sufficient) condition is satisfied
\be      %68
\frac{\partial {\tilde A}^{\pr}}{\partial t} - i[{\tilde A}^{\pr},H]
= 0\,,
\ee
where
$${\tilde A}^{\pr}=fA^\pr+g,\quad A^\pr\equiv uA+vA^{\dg},
$$
$f$ and $g$ (and $u$ and $v$) being functions of time, $z=(z_0-g)/f$,
$f\neq 0$. Indeed, let $|z_0,u_0,v_0\ra$ be an initial SIS, that is an
eigenstate of $A_0=u_0A+v_0A^\dg$. Then the time evolved state
$U(t)|z_0,u_0,v_0\ra$ is eigenstate of operator $A_{{\rm
inv}}=U(t)A_0U^\dg(t)$ with the same eigenvalue $z_0$ ($U(t)$ is the
evolution operator). This operator $A_{{\rm inv}}$ is an integral of
motion\cite{{MMT},{D75}} and satisfy the equation $\partial A_{{\rm
inv}}/\partial t - i[A_{{\rm inv}},H] = 0$. If now $A_{{\rm inv}}$ takes
the form $A_{{\rm inv}}=fA^\pr+g$ we get eq.(68) and the evolved state
is SIS with $z=(z_0-g)/f$. If the HIS $|z\ra$ are an overcomplete set, then
one could get that eq.(68) is also necessary.

So for any given system $H$ we have to look for integrals of motion which
are linear combination of $A$ and $A^\dg$. In canonical case of $A=a$,
$A^\dg = a^\dg$ such linear invariants (and their eigenstates as
well) have been constructed in ref.\cite{MMT} for $n$ dimensional quadratic
Hamiltonians. Here we have to look for systems $H$ which admit integrals
of motion of the form $f\,(ua^2+va^{\dg 2})+g$ to establish stable
evolution and possible generation of s.\,a. SIS $|z,u,v\ra_{sa}$.

The simplest but very important system is the free electromagnetic field
(or equivalently the harmonic oscillator) with $H= \omega (a^\dg a +
1/2) \equiv H_{ho}$ ($\hbar = 1$).  From (68) we get the equations for the
state parameters
\be      %69
\dot{u}+\frac{\dot{f}}{f}u - 2i\omega u = 0,\quad \dot{v}+\frac{\dot{f}}{f}v
+ 2i\omega v= 0,\quad \dot{g}=0\,.
\ee
We easily find solutions
\begin{eqnarray}      %70
g=g_0\equiv const.,\quad f=f_0e^{i\phi_0 t},\quad
z=\frac{z_0-g_0}{f}\nn\\
u=u_0e^{i(2\omega-\phi_0)t},\quad v=v_0e^{-i(2\omega+\phi_0)t},
\end{eqnarray}
where $\phi_0$ is an arbitrary parameter, and $g_0,\, f_0,\, u_0,\, v_0$
are initial values. Note that s.\,a. SIS depend effectively on the ratios
$z/\sqrt{-uv}$ and $v/u$ only (see eqs. (41,42)) which do not depend on
$\phi_0$. Furthermore we take $g_0=z_0-z_0^2$, $f_0=z_0$ and
$\phi_0=-2\omega$ and get $z=z_0\exp(i2\omega t)$.

Thus for the free field Hamiltonian all s.\,a. SIS are stable in time with
parameters
$$\qquad \qquad z=z_0\exp(i2\omega t),\quad u=u_0\exp(i4\omega t),\quad v=v_0
\qquad \qquad \qquad  (70a).$$
This
means that in principle all s.\,a. SIS are realizable for the
electromagnetic field (or harmonic oscillator). It is a separate problem
how to prepare the system in this states or how to generate them from other
known states. As the solutions (70) reveals, the free field evolution operator
$U_{ho}(t)=T\exp(\int H_{ho}(t)dt)$ acts in a highly reducible way on the
set of SIS: it can not change the modulus of $z, u, v$, therefor can't
generate SIS $|z,u,v\ra$ from equal uncertainty HIS $|z\ra$. That is
$U_{ho}$ can't generate s.\,a. squeezing as one expects: the (squared)
variances of s.\,a.  quadratures $X_{sa}, Y_{sa}$ can only oscillate in time
between their minimal and maximal values.  The relative variances of
$X_{sa}, Y_{sa}$, oscillate in time with frequency $4\omega$
($ \psi_0 = {\rm arg}u_0 + {\rm arg}v_0$),
\be      %71
r_{X,Y}(t) \equiv
\frac{\sigma^2_{X,Y}(t)}{\frac{1}{2}|\la[X_{sa},Y_{sa}]\ra|} = |u_0|^2
+|v_0|^2 \mp 2|u_0||v_0|\cos(4\omega t+ \psi_0),
\ee
Let us recall that the relative $q$ and $p$ variances (the ratios
$r_{q,p}(t)$) in the free field evolution of canonical SS oscillate  with
$2\omega$. It worth noting the periodicity of the time evolution of s.\,s.
SIS: the states return their shape after time $T=2\pi/4\omega$ as it is
seen from eqs.(70a) and (41,42).

 We can consider s.\,a. uncertainty ellipses with pulsating in time
semiaxes $a(t)=r_{X}(t)$ and $b(t)=r_{Y}(t)$. Combining this with the
solution (70a) for $z(t)$ we get the picture, quite similar to the known
one for the canonical SS\cite{{Walls},{Nieto}}: the radius
$|z|$ rotates with $2\omega$, the length of semiaxes oscillates with
$4\omega$ (in canonical case the frequencies are $\omega$ and $2\omega$).
These picture applies also to the evolution of the uncertainty ellipses of
$K_1,\, K_2$ in $SU(1,1)$ SIS\cite{D94}, governed by Hamiltonian $H =
2\omega K_3$, where $K_i$ are $SU(1,1)$ generators in any discrete series
representation.

The next Hamiltonian system we consider is general quadratic boson system
(homogeneous for simplicity),
\be      %72
H = \sum_{j=1}^{3}\zeta_{j}(t)K_{j} \equiv H_{quad},
\ee
where the generators $K_{j}$ are quadratic combinations of $a$ and
$a^\dg$, eq.(31).  The necessary and sufficient condition (68) for {\rm
all} s.\,a.  SIS to be stable in time now is not satisfied by $H_{quad}$
unless $\zeta_{1} = 0 = \zeta_{2}$ (the previous case). Then we have to
look for stable evolution of some subsets (which are not overcomplete in
the hole Hilbert space) or to look for other states $|\psi_0\ra$ which
evolve in time as s.\,a. SIS. Let $|\psi_0\ra$ be a state which at $t>0$
(driven by $H_{quad}$) evolves into the set of s.\,a. SIS $|z,u,v\ra_{sa}$,
\be      %73
|z,u,v\ra_{sa}=U(t)|\psi_0\ra,
\ee
where $U(t)$ is the evolution operator, corresponding to $H$. One can
readily see that $|\psi_0\ra$ has to be an eigenstate of
$U^{\dg}(t)A_{sa}(u,v)U(t)$,
\be      %74
U^{\dg}(t)A_{sa}(u,v)U(t)|\psi_0\ra = z|\psi_0\ra.
\ee
When $H=H_{quad}$ the operator $U(t)$ is an element $U_{Mp}$ of the
methaplectic group, which covers the $SU(1,1)$\cite{Perelomov} - the Lie
algebra of the two groups is the same $su(1,1)$. It can be decomposed as
\cite{Perelomov} $U_{Mp}=S(\zeta)\exp(i\tau K_3)$, where $S(\zeta)$ is the
canonical squeeze operator (8). The factor $\exp(i\tau K_3)$ describes the
previous case and for simplicity is omitted henceforth ($\tau=0$). Using
the known BCH formula we rewrite (74) as
\be      %75
(h_3\,K_3 + h_{+}\,K_+ +h_{-}\,K_- )|\psi_0\ra \, = \, z|\psi_0\ra,
\ee
where ($\zeta = re^{i\theta}$)
\begin{eqnarray}      %76
h_3 = -(u+v)\cos\theta\,\sinh(2r),\quad
h_{-}=(u\cosh^2r + v\sinh^2r\,e^{2i\theta},\nn \\
h_{+}=u\sinh^2re^{-2i\theta} + v\cosh^2r.
\end{eqnarray}
We arrived at the conclusion that s.\,a. SIS can be generated from a state
$|\psi_0\ra$ by means of methaplectic evolution operator $U_{Mp}(t)$ iff
$|\psi_0\ra$ is an eigenstate of a complex linear combination (75) of all
$SU(1,1)$ generators (31).

Complex linear combinations of generators of any Lie algebra close another
(larger) algebra, called complex form of the original one. The complex form
of $su(1,1)$ is denoted as $su^c(1,1)$. By this reason we could call the
eigenstates of operators, which are elements of $su^c(1,1)$  {\it algebraic}
$su^c(1,1)$ CS.  S.a. SIS are their particular cases and can be generated
from states $|\psi_0\ra$ from another subset of $su^c(1,1)$ CS, determined by
the relations (75). Solutions of the eq.(75) do exist. Using normal ordered
form of $S(\zeta)$\cite{Perelomov}, the BCH formula and expression (41)  we
can represent the even SIS $|z,u,v;+\ra$ in the form  $|z,u,v;+\ra =
S(\zeta)|\psi_0\ra$ with the following $|\psi_0\ra$ ($\zeta =
re^{i\theta}$),
\be      %77
|\psi_0\ra  = {\cal N}_{+}\cosh^{-2}|\xi|\,M\l(a_+,\frac{1}{2};p_{2}
(a,a^\dg)\r)|0\ra \equiv |z,\xi,\phi_{u};+\ra_{Kummer},
\ee
where $\phi_{u}$ is the phase of $u$, $\xi = \tanh
r\exp(-i\theta)=\sqrt{-v/u}$,\, $|\xi|<1 $,\,  $M(a,b,z) \equiv\,
_{1}F_{1}(a,b;z)$ is the Kummer function \cite{Stegun}, $a_{+} =
(1+z/\sqrt{-uv})/4$ and $p_{2}(a,a^\dg)$ is a second order polynomial
of $a$ and $a^\dg$ (element of $su(2,C)$),

$$p_2(a,a^\dg) =
\xi\,\ln(1-|\xi|^2)\l(a^{\dg 2}+\xi^2a^2 - \xi\,a^{\dg}a -
\xi/2\r),$$

One can consider $z$, $\xi$ and the angle $\phi_{u}$ as free parameters of
an initial $su(2,C)$ CS which evolves into SIS $|z,u,v;+\ra$.
Similarly one can treat the problem of methaplectic generation of odd s.\,a.
SIS $|z,u,v;-\ra$.

Consider some particular cases of the Kummer function states (77). When
$a_{+}(z,\xi,\phi_u)$ $= 0$ (then $M(0,b,z)=1$) we get the initial states
as $|0\ra$ and then s.\,a. SIS $|z,u,v;+\ra$ are Perelomov CS (canonically
squeezed vacuum). When $a_{+}(z,\xi,\phi_u)= 1/2$ the Kummer function is
an exponent and the initial state is a Perelomov CS, the final one being
again such CS and s.\,a. SIS as well.  Perelomov CS are stable under the
action of methaplectic evolution operators.  When
$a_{+}(z,\xi,\phi_u)=-m=-1,-2,...$ the above Kummer function is a Hermite
polynomial\cite{Stegun} $H_{2m}(2\sqrt{p_{2}(a,a^\dg)})$ and the
initial state is clearly a finite superposition of Fock states
$|n\ra$. Following ref.\cite{BHYu} we can call
$H_{2m}(2\sqrt{p_{2}(a,a^\dg)})|0\ra$ Hermite polynomial states -
these are more general and recover those in \cite{BHYu} when
$p_2(a,a^\dg)={\rm const}.a^{\dg 2}$.  The squeeze operator
$S(\zeta)$ does not affect the Kummer function parameter $a_{+}$.
Therefor in this case the final state is an even s.\,a. SIS of the form
$S(\zeta)H_{2m}(\sqrt{\xi} a^{\dg})|0\ra$.

Fock states $|n\ra$ can be constructed by means of $n+1$ Glauber CS
$|\alpha\ra$ and in principle any finite superposition of $|n\ra$ as
well\cite{Yansky}. Thus the subset of s.\,a. SIS of the form of squeezed
Hermite polynomial states can be in principle experimentally created,
using Hermite polynomial states e.g. as input states for the degenerate
parametric amplifier. To get other type of s.\,a. SIS by means of
methaplectic evolution operator one need first to create Kummer function
states, which in general are infinite superpositions of number states.
Due to the fact that s.\,a. HIS $|z\ra_{sa}$ are available ( e.g. the
e.\,o. CS $|\alpha_{\pm}\ra$\cite{GGKnight} or the Yurke-Stoler
states\cite{YS}) it is desirable to look for generation of s.\,a. SIS from
s.\,a. HIS. However the s.\,a. squeeze operator $S_{sa}(u,v)$ is not
unitary (it is isometric only) and one has to look for processes with
nonunitary quantum dynamic of boson system. In principle this could be
expected if interaction with other systems is present not in parametric
form.

Dealing with s.\,a. squeezing by methaplectic evolution
it is natural to try to produce s.\,a. SS other than
SIS, taking as input some of experimentally available boson states. As such
input states let us consider the Glauber CS $|\alpha\ra$, the Fock states
$|n\ra$ and the e.\,o. CS $|\alpha_{\pm}\ra$. After
some standard but long calculations and analysis we can find that quadratic
squeezing (after the definition (3)) occurs in the third case only. As an
example we take the ordinary squeezed even states $|\zeta,z;+\ra$,
\be       %78
|\zeta,z;+\ra = \exp\l[\frac{1}{2}(\zeta a^{\dg
2}-\zeta^*a^2)\r]|z;+\ra,
\ee
where $|z;+\ra = |\alpha_+\ra, z=\alpha^2$.
The variances of the amplitude quadrature $q$ and s.\,a.
quadrature $X_{sa}$ and the Mandel Q-factor in $|\zeta,z;+\ra$ are obtained
in the form ($\zeta=r\,e^{i\theta}$, $z=\rho e^{i\phi}$)
\begin{eqnarray}      %79
\Delta_q^2(\zeta,z) = \frac{1}{2}+\sinh^2r+\la a^\dg
a\ra\l(\cosh(2r)+\cos\theta\,\sinh(2r)\r)+\frac{1}{2}\cos\theta\,
\sinh(2r)\qquad \nn \\
+{\rm Re}\l[z\l(\cosh r+e^{-i\theta}\sinh r\r)^2\r],\quad\\ %79
\Delta_{X_{sa}}^2(\zeta,z) =
2\cos^2\theta\,\sinh^2(2r)\l[\rho^2+\la a^\dg
a\ra\l(1-\la a^\dg a\r)\r] +
\l(s_1^2+s_2^2+2s_1s_2\cos(2\theta)\r)\quad \nn \\
\times\l(1+2\la a^\dg a\ra\r)+
4\rho\cos\theta\,\sinh(2r)\l[s_1\cos(\phi)+s_2\cos(2\theta-\phi)\r],
\quad \\ %80
\la+;z,\zeta|a^\dg a|\zeta,z;+\ra Q(\zeta,z) =
\cosh^2(2r)\l[\rho^2+\la n\ra\l(1-\la n\ra\r)\r]
+\frac{1}{2}\sinh^2(2r)\l(1+2\la n\ra\r)\quad \nn \\
-\la n\ra\cosh(2r) - \sinh^2r +
\rho\sinh(2r)\cos(\theta-\phi) \l[2\cosh(2r)-1\r],\quad  %81
\end{eqnarray}
where
$\la a^\dg a\ra\equiv \la+;z|a^\dg a|z;+\ra,\quad
s_1=\cosh^2 r,\,\, s_2=\sinh^2 r.$

Strong $q$-squeezing (asymptotically, when $r \rightarrow \infty $, absolute
one) we get, e.g. in the family of states $|\zeta=-r,z=-\rho;+\ra$.
Quadratic squeezing ($X_{sa}$-squeezing) is obtained e.g. in the family
$|\zeta=r,z=-\rho;+\ra$ with small $r$ and $\rho$. This is illustrated by
the plots on Fig.4. Since the quadratic squeezing here is not strong and is
observed in short interval of $r$ we have used scaling factors in order to
combine the two graphics in one figure. Joint $q$ and $X_{sa}$ squeezing is
also possible - it occurs e.g. in $|\zeta\!=\!r,z\!=\!-0.4;+\ra$ when
$0.12\!<\!r\!<\!0.34$. The photon statistics in these linear and quadratic
amplitude SS is obtained as superpoissonian.  Subpoissonian statistics
occurs in the families $|\zeta=\pm ir,z=\mp i\rho;+\ra$ for small $r$ and
large $\rho$, e.g. for $r=0.1$ and $\rho>2$ (but they again are not SS).
The occurrence of linear amplitude squeezing in $|\zeta,z;+\ra$ is normal
since application of the canonical $q$-$p$ squeeze operator $S(\zeta)$ to
any ($\zeta$ independent) state always produce $q$- and $p$-squeezing.  This
result stems from transformation properties of the uncertainty matrix
$\sigma(\rho;q,p)$ under linear canonical transformations (see section
III)\cite{D95}.  Generation of quadratic squeezing by means of $S(\zeta)$
was not quite expected.  We note that s.\,a. SS $|\zeta,z;+\ra$ can be
easily realized since HIS $|z;+\ra$ are available and could be used e.g. as
input states in degenerate parametric amplifier.

\bigskip
\bc
{\large{\bf VI. Conclusion }}
\ec
\medskip
We have considered some general properties of states $|z,u,v\ra$, which
minimize the Schr\"odinger uncertainty relation (5) for arbitrary pair of
observables $X$ and $Y$, terming such states Schr\"odinger intelligent
states (SIS). SIS are eigenstates of complex linear combination of $X$ and
$Y$. The uncertainty matrix for $X$ and $Y$ in any state with density
matrix $\rho$ can be diagonalized by linear transformation of $X$ and $Y$,
which preserves the commutator $[X^\pr,Y^\pr] = [X,Y]$. Such
transformation is an $SU(1,1)$ transformation and when $[X,Y] = i$ it is
the canonical one. In the important physical case of $X, Y$ being the
quadratures of the squared boson (photon) destruction operator $a^2$ all
SIS are explicitly constructed and discussed.

The even and odd s.\,a.  SIS contain in a natural way many known
Schr\"odinger cat states and exhibit interesting physical properties such
as very strong amplitude and squared amplitude squeezing (even
simultaneously), super- and subpoissonian statistics. A subset of s.\,a.
which are of the form of ordinary squeezed complex Hermite polynomial
states could be realized in the degenerate parametric amplifier scheme
using the finite (Hermite polynomial) superposition of number states as an
input.  Number states can be in principle experimentally
created\cite{Yansky}.  We have also shown that the ordinary squeezed even
CS, $S(\zeta)|z;+\ra=|\zeta,z;+\ra$, eq.(78), can exhibit
strong amplitude and light amplitude-squared squeezing.
These double squeezed states, $|\zeta,z;\pm\ra$,
can be easily realized e.g. in the degenerate amplifier scheme, since the
ordinary even and odd CS are available\cite{{Hach},{GGKnight}}. It is
desirable from this point of view to examine for linear and quadratic
squeezing other ordinary squeezed equal uncertainty HIS
$S(\zeta)|z\ra_{sa}$ (in a recent paper [33] such states have been briefly
considered as "$SU(1,1)$ minimum uncertainty states with equal variance in
two observables").

From the algebraic point of view s.\,a. SIS and other s.\,a. states
considered here are eigenstates of operators, which are complex linear
combinations of the $SU(1,1)$ generators $K_i$ in the s.\,a.
representation (31). The set of all complex linear combinations of $K_i$
closes the Lie algebra $su^c(1,1) \sim sl(2,C)$, which is the complex form
of $su(1,1)$. By this reason one can call such eigenstates $su^c(1,1)$ CS.
So we have constructed here several subsets of the $su^c(1,1)$ CS in the
representation (31). In this representation an other subset of $su^c(1,1)$
CS (different from ours $|z,u,v\ra_{sa}$, $|\zeta,z;\pm\ra$ and Kummer
function states (77)) has been constructed in the recent paper by
W\"unsche\cite{Wunsche}. In ref.\cite{D94} eigenstates of $uK_1+vK_2$ (the
$SU(1,1)$ SIS) for any discrete series (square integrable) representation
have been constructed using Barut--Girardello CS representation.
Eigenstates of $K_3\pm iK_1$ are considered in the very recent
paper\cite{Puri}. Let us note that the ordinary squeezed Fock states
$S(\zeta)|n\ra$ are also s.\,a. $su^c(1,1)$ CS.

After the first e-print submission my attention was kindly brought to the
recent preprints \cite{Brif} where eigenstates of complex linear
combinations of $SU(1,1)$ generators are also constructed (using the
similar approach) and discussed as squeezed and intelligent states.
Eigenstates of complex combinations $ua^2 + va^{\dagger 2}$ are also
presented in \cite{D95} (using present approach) and in \cite{Shanta}
(using an algebraic approach).
\medskip

{\bf Acknowledgment.} The work is partially  supported  by  the  Bulgarian
science foundation under contract \# F-559.
\bigskip
\bc
{\large {\bf VII. Appendix}}
\ec
\medskip

{\bf Nonunitarity of the squared amplitude squeeze operator.} The
squeeze operator for arbitrary pair of observables $X$, $Y$ is defined as
a map "HIS $\rightarrow$ SIS",
\be      %82
 |z\ra \,\,\longrightarrow \,\, |z,u,v\ra = S(u,v)|z\ra
 \ee
This defines the operator $S(u,v)$ correctly in the hole Hilbert space if
the set of HIS $|z\ra$ is overcomplete\cite{Klauder}, since in such cases
any state $|\psi\ra$ is linear span of HIS $|z\ra$. Such is the case of
squared amplitude (s.\,a.) HIS and SIS, we are interesting here. Indeed, any
$|\psi\ra$ is known to be decomposed as a sum of one even and one
odd state, $|\psi\ra = |\psi_{+}\ra + |\psi_{-}\ra$ and every
$|\psi_{\pm}\ra$ can be expanded as an integral over
even/odd CS $|\alpha_{\pm}\ra$ using the resolution formulas (23).
For breavity here we omit the subscrip {\small sa} in s.\,a. SIS
$|z,u,v\ra$ and s.\,a. squeeze operator.

The so defined s.\,a. squeeze operator $S(u,v)$ is isometric, i.e. it
preserves the norm of the states. This can be proved most easily using the
analyticity of the canonical CS representation\cite{Klauder}: the canonical
CS diagonal
matrix elements $\la\alpha|Z|\alpha\ra$ uniquely determine the operator
$Z$. We take $Z=S^\dg S$ and consider
$\la\alpha|S^{\dg}S|\alpha\ra$. The operator $S$ (and $S^\dg S$ as
well) by definition preserves the parity of the states. Then expressing CS
$|\alpha\ra$ in terms of even and odd CS $|\alpha_\pm\ra \equiv |z;\pm\ra,
z=\alpha^2$, $$|\alpha\ra = f_+(\alpha)|\alpha_+\ra +
f_-(\alpha)|\alpha_-\ra,$$ where $|f_+|^2 + |f_-|^2 = 1$, and taking into
account that the constructed even/odd SIS $|z,u,v;\pm\ra=S|z;\pm\ra$ are
normalized and orthogonal to each other, we arive at
$\la\alpha|S^{\dg}S|\alpha\ra = 1$, which proves that $S(u,v)$ is
isometric, $S^\dg S = 1$.

Now we shall prove that s.\,a. $S(u,v)$ is not unitary, that is $SS^\dg
\neq 1$. The proof can be carred out by admitting the inverse. Let
$SS^\dg = 1$. Considering the canonical CS diagonal matrix elements of
$A=a^2$ and $S^\dg A^\pr S$ ($A^\pr = uA+vA^\dg$) we get the
same result, i.e.  $A=S^\dg A^\pr S$ and then $A^\pr =
SAS^\dg$. Now let us recall that the commutators
$[A^\pr,A^{\pr\dg}]$ and $[A,A^\dg]$ are equal as a
consequence of $|u|^2-|v|^2=1$ (othervise they are proportional). Thus if
$S$ is unitary then it commutes with the operator $[A,A^\dg]$ and then
the mean commutator $\la v,u,z|[A,A^\dg]|z,u,v\ra$ in SIS $|z,u,v\ra$
does not depend on parameters $u$ and $v$. But if
$\la v,u,z|[A,A^\dg]|z,u,v\ra = \la z|[A,A^\dg]|z\ra$
then from the formula (54), (55) we could get negative variances of $q$ and
$p$ for fixed $z$ and large $|v|$. Indeed, ${\rm Re}[(u-v)z^*] =
|z(u-v)|\cos(\phi_1)$ and $|u-v|^2=1+2|v|^2-2|uv|\cos(\phi_2)$,
where $\phi_1 = \phi_2 - {\rm arg}z$. Then for e.g. $\phi_2 = \pi/2$,
${\rm arg}z = -\pi/2$ and large $|v|$ we get ${\rm Re}[(u-v)z^*] =
-|z|(1+2|v|^2)$ which leads to negative variance of $q$, eq.(54), for large
$|v|$. End of the proof.

It worth to note that the set of operators $S(u,v)$ realizes a nonunitary
representation of $SU(1,1)$. Indeed, one can check that the product of two
commutator preserving $SU(1,1)$ transformations (13) is again such a
transformation and the inverse as well. Thus both the amplitude (the
ordinary) squeeze operators (we mean the full methaplectic operators (33),
$U(\vec{\zeta})=S(\zeta)\exp(i\tau K_3)$) and the squared amplitude squeeze
operators realize representations of $SU(1,1)$ - in the first case it is
unitary, in the second case it is isometric only. However so far we do not
have the squared amplitude squeeze operator expreesed in closed form in
terms of boson operators $a$ and $a^\dg$. We note that one can apply to
s.\,a. HIS $|z\ra_{sa}$ the metaplectic unitary operators $U(\vec{\zeta})$
(in particular $S(\zeta)$) and obtain an other large family of
states $|\vec{\zeta},z\ra_{sa}$, which however are not $K_1$-$K_2$ SIS.

\newpage

\bc       {\bf F i g u r e \hspace{5mm} c a p t i o n s  }  \ec
\vspace{5mm}
\bc
Fig.1. {\small Variance $\Delta^2 q(x)$ in even SIS
$|z,u,v;+\rangle$,\, $u=\sqrt{1+|v|^2}$.\,\,\, a) $z=-1$,\\
$v=-x$;$\quad$ b) $z=-2,\,v=-x$;$\quad$ c) $z=-5,\, v=-x,\,\,\, x>0$.\\
The variance is squeezed if $\Delta^2 q<0.5$. }  \ec
\vspace{3mm}
\bc
Fig.2. {\small Variance $\Delta^2 Y_{sa}(x)$ in even SIS $|z,u,v;+\rangle$.
 \,\,a, b, c\, the same\\
states as in Fig.1. The s.\,a. variance is squeezed if $\Delta^2 Y_{sa}<1$.}
\ec
\vspace{3mm}
\bc
Fig.3. {\small Photon distribution $f(n)$ in even SIS
$|z,u,v;+\rangle$, $u=\sqrt{1+|v|^2}$.\\ $\quad\qquad$ a)
$z=-5,\, v=-6$:\, $Q>0$, double SS;\,\, b) $z=-5,\,v=0.2$:\,
$Q<0$, not SS}. \ec
\vspace{3mm}

\bc
Fig.4. {\small Linear and quadratic
squeezing in even states
$|\zeta,z;+\rangle$,\,\,  eq. (78).} \\
{\small$\qquad$ a) $\zeta=-r,\, z=-4$,\, $f(r)=2\Delta^2q(r)$.} \\
{\small$\qquad$ b) $\zeta = 0.3r$, $z = -0.4$,\,
$f(r) = \Delta^2 X_{sa}(0.3r)$}. \ec

\end{document}